# Comment on "Observation of Shapiro Steps in the Charge Density Wave State Induced by Strain on a Piezoelectric Substrate"

D.Yu. Saltykova[1,2], M.V. Nikitin[1], V.Ya. Pokrovskii[1], S.G. Zybtsev[1], V.V. Kolesov[1], V.V. Kashin[1], I.E. Kuznetsova[1], I.A. Nedospasov[1]

[1]Kotelnikov Institute of Radioengineering and Electronics of RAS, Mokhovaya 11-7, 125009 Moscow, Russia

[2]Moscow Institute of Physics and Technology (National Research University), 141700 Dolgoprudnyi, Moscow oblast, Russia

In their Letter Fujiwara *et al.* [1] report a high-quality experiment demonstrating the synchronization of the CDW sliding in $NbSe_3$ whiskers (nanowires) with surface acoustic waves (SAWs). The SAWs are induced in the conventional $LiNbO_3$ piezoelectric substrates through application of rf voltage to an interdigital transducer (IDT). When a SAW mode is excited, Shapiro steps (ShSs) appear on the I-V curves, while no ShSs are observed when the frequency is shifted from the resonance. This gives clear evidence of synchronization of the CDW with the acoustic modes.

The authors also argue that the ShSs are induced by the strain field of the wave. In fact, the comparison of the ShSs induced by SAWs and by rf voltage applied directly to the whiskers, as well as of their amplitude dependences, reveals qualitative differences. However, we suggest that the authors might have ignored the spatial heterogeneity of the field of the wave, whose wavelength, $\lambda = 13.2$ μm, is comparable with the distances between the current and potential probes for all the structures reported. We have performed similar studies on another CDW compound and in another frequency range with different acoustic wave types and got similar results. However, the differences between the wave- and rf-field-induced ShSs vanish if $L<<\lambda$.

We had put samples of $NbS_3$ (at 300K) and $TaS_3$ (at ~120K) on different piezoelectric devices and observed ShSs on the I-V curves in the resonance modes [2,3,4,5]. Further we chose $TaS_3$ as the compound with more pronounced effects of CDW-lattice coupling [6,7]. The samples were arranged on a 350 µm thick YX $LiNbO_3$ delay line. We excited plate acoustic modes with the frequencies of 1.131 MHz ($SH_0$ wave), 6.02 MHz ($S_1$ Lamb wave), 9.361 MHz ($SH_1$ wave) and 16.685 MHz ($S_3$ Lamb wave).

Fig.1a shows current ($I$) dependences of differential resistance ($R_d$) with the 6.02 MHz rf voltage applied to the sample ($V_s$) and to the IDT ($V_{IDT}$). One can see that the ShSs induced by the wave are wider. The decrease of the width of the 0-th ShSs [8] vs. $V_{rms}$ (inset) is more gradual, as well as of the 1$^{st}$ ShS studied at 1.131 MHz (Fig.1c), like in [1].

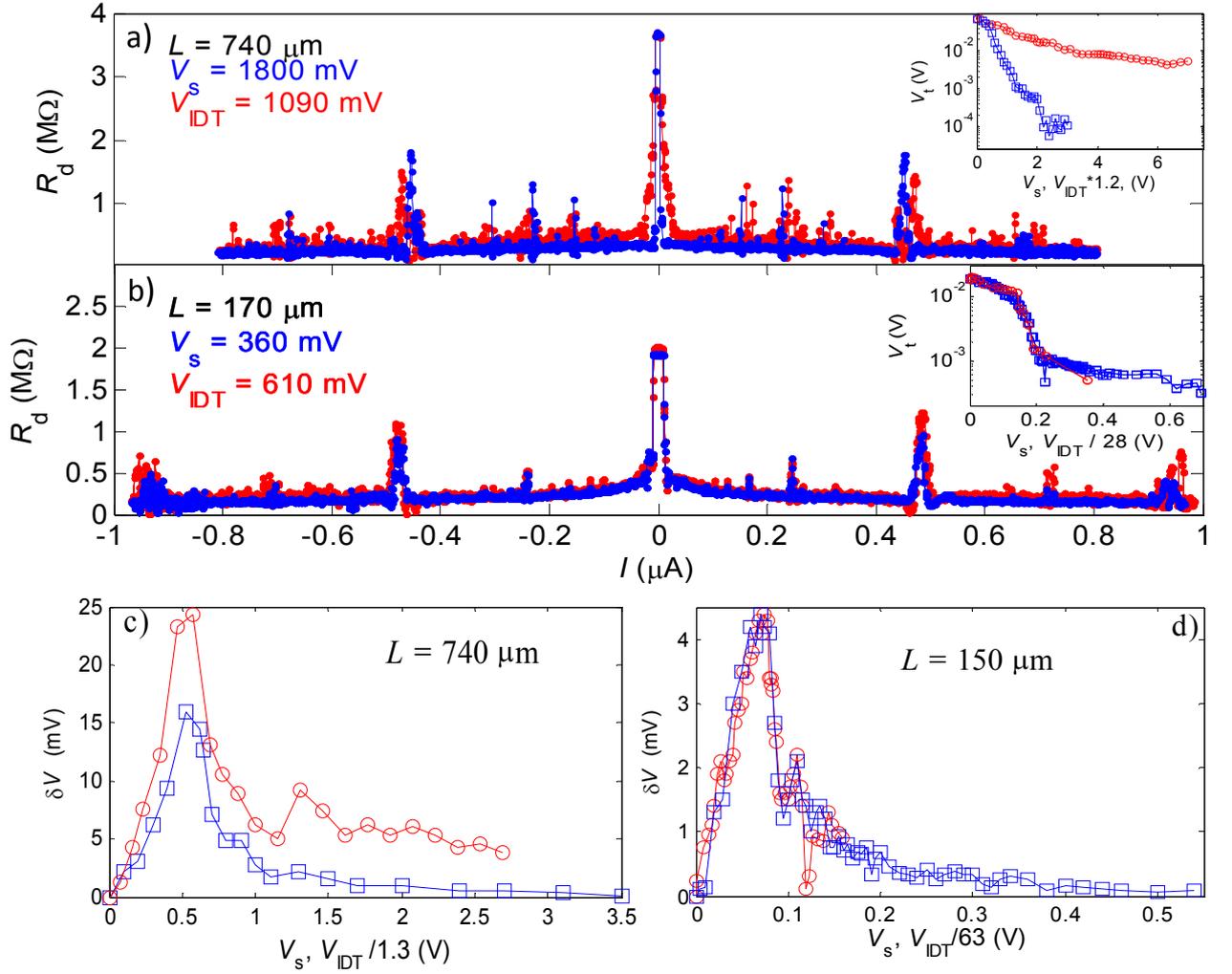

FIG.1. $R_d(I)$ for the 740 μm-long sample (a) and for the 170 μm-long segment (b) under the impacts at 6.02 MHz. Insets: the corresponding amplitude dependences of the threshold voltage [8]. c) and d) show the amplitude dependences of the 1$^{st}$ ShS widths under the impacts at 1.131 MHz for the the 740 μm-long and 150 μm-long segments. Blue (red) curves: rf voltage applied to the sample (IDT).

In the sample studied the contacts were separated by $L=740$ μm, while $\lambda=3$ mm ($L\approx1/4\ \lambda$). Afterwards two additional contacts were deposited to form 3 segments of $L=200$, 170 and 150 μm length to approach the case of $L<<\lambda$. The differences between the two types of ShSs have virtually disappeared (Fig. 1b), as well as between their amplitude dependences (compare insets to Fig.1a,b for the 0-th ShSs and Figs.1c,d for the 1$^{st}$ ones).

Similar features of the ShSs on the $R_d(I)$ curves and their modifications at the short segment were observed at 2 other plate wave modes.

Thus, one should carefully consider the effects of non-uniform field of an acoustic wave on the CDW. If $L\sim\lambda$, at certain time intervals the field applies large opposite forces in different parts of the nanowire, and the CDW can be broken into domains sliding at different velocities and synchronized at different currents [1]. This conclusion can concern both electric and elastic fields of the wave. It can concern also the results of [9,10]: although the mechanical origin of the ShSs has been proven, the reported differences between them and electric-field induced ShSs might be attributed to inhomogeneous strains at the vibration resonances.

On the other hand, given the "horizontal" effect of strain [11], it seems likely that the "mechanical" ShSs show features common to those of electrical ShSs, e.g., oscillations of their magnitudes [1]. In the weak-pinning model a strain of a whisker results into a modification of the periodic pinning potential [9]. Correspondingly, a strain periodic in time results in a periodic change of the potential with respect to the CDW, i.e., a shift of the CDW relative to the periodic potential, like in the case of electric-field induced ShSs.

Acknowledgments—This work was supported by RSF, grant No 25-29-00876

Data availability—The more detailed data are available under a reasonable request.

D.Yu. Saltykova, M.V. Nikitin, V.Ya. Pokrovskii[*], S.G. Zybtsev, V.V. Kolesov, V.V. Kashin, I.E. Kuznetsova, I.A. Nedospasov

Kotelnikov Institute of Radioengineering and Electronics of RAS, 125009 Moscow, Russia

[*]Contact author: vadim.pokrovskiy@mail.ru

Koji Fujiwara, Takuya Kawada, Natsumi Nikaido, Jihoon Park, Nan Jiang, Shintaro Takada, and Yasuhiro Niimi, Observation of Shapiro Steps in the Charge Density Wave State Induced by Strain on a Piezoelectric Substrate, Phys.Rev. Lett. 2025; arXiv:2511.09888 [cond-mat.mes-hall], 10.48550/arXiv.2511.09888. 2025.
1. Koji Fujiwara, Takuya Kawada, Natsumi Nikaido, Jihoon Park, Nan Jiang, Shintaro Takada, and Yasuhiro Niimi, Observation of Shapiro Steps in the Charge Density Wave State Induced by Strain on a Piezoelectric Substrate, Phys.Rev. Lett. 2025; arXiv:2511.09888 [cond-mat.mes-hall], 10.48550/arXiv.2511.09888. 2025.

2. M.V. Nikitin, V.Ya. Pokrovskii, S.G. Zybtsev, V.V. Kolesov, V.V. Kashin, The Effect of High-Frequency Mechanical Vibrations on the Transport Properties of Quasi-One-Dimensional Conductors, in: XV Russian Conference on Semiconductor Physics, Nizhny Novgorod, October 3–7, 2022. Abstracts, p. 398 (in Russian); https://semicond2022.ru/file/29/b86bdcf0/abstracts.pdf.

3. M.V. Nikitin, V.Ya. Pokrovskii, D.A. Kai, S.G. Zybtsev, V.V. Kolesov, V.V. Kashin, Shapiro steps in synchronization of a charge density wave by acoustic waves, XX Conference "Strongly Correlated Electron Systems and Quantum Critical Phenomena", Lebedev Physical Institute, Moscow, May 25, 2023, ABSTRACT COLLECTION, ISBN 978-5-4344-0993-3, M.-Izhevsk: Institute of Computer Research, 2023, p. 138 (in Russian); https://sces.lebedev.ru/wp-content/uploads/2023/07/SCES_A5.pdf.

4. M.V.Nikitin, V.Ya. Pokrovskii, D.A. Kai, S.G. Zybtsev, V.V. Kolesov, V.V. Kashin, Shapiro steps at charge density wave synchronization by electric and acoustic oscillations, Proceedings of the III International Conference "Physics of Condensed Matter" FKS-2023, Edited by Dr. Sci. (Physics and Mathematics) B.B. Straumal. Chernogolovka, May 29 – June 2, 2023, p. 139 (in Russian); http://www.issp.ac.ru/fks2023/assets/files/Abstract_FKS_7.pdf

5. M.V. Nikitin, V.Ya. Pokrovskii, D.A. Kai, S.G. Zybtsev, "Dynamics of the charge density wave under electrical and mechanical vibrations", Proceedings of the XXVIII International Symposium, March 11–15, 2024, Nizhny

Novgorod, IAP RAS, ISBN 978-5-8048-0124-4, v.2, p. 759 (in Russian); https://nanosymp.ru/ru/file/177/987a7142/2024_v2.pdf

6. J. W. Brill, Elastic properties of low-dimensional materials, in Handbook of Elastic Properties of Solids, Liquids, and Gases, edited by M. Levy, H. E. Bass, and R. R. Stern (Academic Press, San Diego, 2001), Vol. 2, p. 143.

7. V. Ya. Pokrovskii, S. G. Zybtsev, M. V. Nikitin, I. G. Gorlova, V. F. Nasretdinova, S. V. Zaitsev-Zotov, High-frequency, `quantum' and electromechanical effects in quasi-one-dimensional charge density wave conductors, Physics-Uspekhi **56**, 29 (2013).

8. The threshold voltage can be considered as the 0-th ShS half-width.

9. M. V. Nikitin, S. G. Zybtsev, V. Ya. Pokrovskii, and B. A. Loginov, Mechanically induced Shapiro steps: Enormous effect of vibrations on the charge-density wave transport, Appl. Phys. Lett. **118**, 223105 (2021).

10. M. V. Nikitin, V. Ya. Pokrovskii, D. A. Kai, and S. G. Zybtsev, On the Fundamental Difference between the Effects of Electrical and Mechanical Vibrations on the Dynamics of a Charge Density Wave, JETP Letters, **118**, 854 (2023). 10.1134/S0021364023603342.

11. Yu Funami and Kazushi Aoyama, Shapiro Steps and Surface Acoustic Waves in Charge Density Wave Dynamics, New Physics: Sae Mulli **73**, 1086 (2023).